\RequirePackage{fix-cm}
\documentclass[smallextended]{svjour3}       
\smartqed  
\usepackage{amsmath}
\usepackage{ams}
\usepackage{graphicx}
\begin{document}

\title{\bf Saari's Conjecture and Variational Minimal Solutions for $N$-Body Problems\thanks{Supported partially by NSF of China}
}


\author{Yu Xiang          \and
        Zhang Shiqing 
}


\institute{College of Mathematics and Yantze Center of Mathematics, Sichuan University, Chengdu 610064,P.R.China\\
              \email{xiang.zhiy@gmail.com}\\
              \email{ zhangshiqing@msn.com}
}

\date{Received: date / Accepted: date}

\maketitle

\begin{abstract}
In this paper, we will prove Saari's conjecture in a particular case by using a arithmetic fact, and then, apply it to prove that for any given positive masses, the variational minimal solutions of the N-body problem in ${\mathbb{R}}^2$ are precisely a relative equilibrium solution whose
configuration minimizes the function $IU^2$ in ${{\mathbb{R}}}^2$.
\end{abstract}

\section{Introduction}
\label{intro}
~~~ In 1970, Donald Saari \cite{saari1970bounded} proposed the following conjecture : {\it In
the Newtonian $N$-body problem, if the moment of inertia,
$I=\Sigma^n_{k=1}m_k|q_k|^2$, is constant, where $q_1,q_2,\cdots
,q_n$ represent the position vectors of the bodies of masses
$m_1,\cdots ,m_n$, then the corresponding solution is a relative
equilibrium.} In other words: Newtonian particle systems of constant
moment of inertia rotate like rigid bodies.

A lot of energies have been spent to understand Saari's conjecture, but
most of those works \cite{palmore1979relative,palmore1981saari} failed to achieve crucial results.
But there have been a few successes in the struggle to understand
Saari's conjecture. McCord \cite{mccord2004saari} proved that the conjecture is true for
three bodies of equal masses. Llibre and Pina \cite{llibre2002saari} gave an
alternative proof of this case, but they never published it.In particular,
Moeckel \cite{moeckel2005computer,moeckel2005proof} obtained a computer-assisted proof for the Newtonian
three-body problem with positive masses when physical space
is $\mathbb{R}^d$ for all positive integer $d\geq2$. Diacu,
P$\acute{\rm e}$rez-Chavela, and Santoprete \cite{diacu2005saari} showed that the
conjectre is true for any $n$ in the collinear case for potentials
that depend only on the mutual distances between point masses.
There have been results, such as \cite{santoprete2004counterexample,roberts2006some,schmah2007saari}, which
studied the conjecture in other contexts than the Newtonian one.

Recently the interest in this conjecture
has grown considerably due to the discovery of the figure eight
solution \cite{chenciner2000remarkable}, which, as numerical arguments show, has an
approximately constant moment of inertia but is not a relative
equilibrium.

The variational minimal solutions of the N-body problem are attractive, since they are nature from the viewpoint of the principle of least action.  Unfortunately, there were very few works about the variational minimal solutions before 2000.
It's worth noticing that a lot of results have been got by the action minimization methods in recent years, please see \cite{barutello2004action,chen2001action,chen2003binary,chen2008existence,Chenciner2002,chenciner2000remarkable,chenciner2000minima,ferrario2004existence,long2000geometric,zhang2002variational,zhang2004nonplanar,zhang2004new,zhang2001minimizing} and the references there.

In this paper, we will first prove a arithmetic fact, then use it to prove Saari's conjecture in a particular case: the position $q_i(t)$ of $i$-th point particle has the form \begin{equation}
q_i(t)=a_i\cos(\frac{2\pi}{T}t)+b_i\sin(\frac{2\pi}{T}t),~~~~~~\forall t\in\mathbb{T}.
\end{equation}
and $a_i, b_i\in\mathbb{R}^d$ for all $i=1,\ldots, N $. In the last part, we describe the shapes of the variational minimal solution of the N-body problem in some constraints.

Let $\mathcal{X}_d$ denote the space of configurations of $N\geq 2$ point particles with masses $m_1, \ldots,m_N$ in Euclidean space $\mathbb{R}^d$ of dimension $d$, whose center of masses is at the origin, that is, $\mathcal{X}_d = \{ q = (q_1,\cdots, q_N)\in (\mathbb{R}^d)^N: \sum_{i = 1}^{N} {m_i q_i} = 0  \}$. Let $\mathbb{T} = \mathbb{R}/T\mathbb{Z} $ denote the circle of length $T = |\mathbb{T}|$, embedded as $\mathbb{T} \subset \mathbb{R}^2$.By the loop space $\Lambda$, we mean the Sobolev space $\Lambda  = H^1(\mathbb{T}, \mathcal{X}_d)$. We consider the opposite of the potential energy defined by
\begin{equation}
U(q) = \sum_{i<j} {\frac{m_i m_j }{|q_i - q_j|}}.
\end{equation}
The kinetic energy is defined (on the tangent bundle of $\mathcal{X}_d$) by $K = \sum_{i=1}^{N} {\frac{1}{2}{m_i |\dot{q}_i|^2}}$, the total energy is $E = K- U$ and the Lagrangian is $L(q,\dot{q}) = L = K + U = \sum_i \frac{1}{2} m_i |\dot{q}|^2  + \sum_{i<j}{\frac{m_i m_j}{|q_i - q_j|}}$.
Given the Lagrangian L, the positive definite functional $\mathcal{{A}}:\Lambda \rightarrow \mathbb{R} \cup \{+\infty\}$
defined by
\begin{equation}
\mathcal{{A}}(q) = \int_{\mathbb{T}}{ L(q(t),\dot{q}(t)) dt}.
\end{equation}
 is termed action functional (or the Lagrangian action).

The action functional $\mathcal{{A}}$ is of class $C^1$ on the subspace $\hat{\Lambda} \subset \Lambda $, which is collision-free space. Hence critical point of $\mathcal{{A}}$ in $\hat{\Lambda}$ are T-periodic classical solutions (of class $C^2$) of  Newton's equations
\begin{equation}
m_i \ddot{q}_i = \frac{\partial U}{\partial q_i}.
\end{equation}

{\bf Definition \cite{chenciner2000remarkable}.} A configuration $q=(q_1,\cdots,q_N)\in {\mathcal{X}}_d\setminus\Delta_d$ is called a central configuration if there exists a constant $\lambda\in {\mathbb{R}}$ such that
\begin{equation}
\sum_{j=1,j\neq k}^N \frac{m_jm_k}{|q_j-q_k|^3}(q_j-q_k)=-\lambda m_kq_k,1\leq k\leq N
\end{equation}

The value of $\lambda$ in (1.1) is uniquely determined by
\begin{equation}
\lambda=\frac{U(q)}{I(q)}
\end{equation}

Where
\begin{equation}
\Delta_d=\left\{q=(q_1,\cdots,q_N)\in (\mathbb{R}^d)^N: q_j=q_k~\mbox{for~some}~j\neq k\right\}
\end{equation}
\begin{equation}
I(q)=\sum_{1\leq j\leq N} m_j|q_j|^2~~~~~~~~~~~~~~~~~~~
\end{equation}

It's well known that the central configurations are the critical points of the function $IU^2$, and $IU^2$ attains its infimum on ${\mathcal{X}}_d\setminus\Delta_d$. Furthermore, we know \cite{moeckel1990central} that $inf_{{\mathcal{X}}_2\setminus\Delta_2}{IU^2}<inf_{{\mathcal{X}}_1\setminus\Delta_1}{{IU^2}}$and $inf_{{\mathcal{X}}_3\setminus\Delta_3}{{IU^2}}<inf_{{\mathcal{X}}_2\setminus\Delta_2}{IU^2}$ when$N\geq4$. It is well known that the homographic solutions derived by the central configurations which minimize the function $IU^2$ when $N\geq4$ and ${\mathbb{R}^d}={\mathbb{R}^3}$ are homothetic, furthermore, a homographic motion in ${\mathbb{R}^3}$ which is not homothetic takes place in a fixed plane\cite{albouy1997probleme,arnold2006dynamical,Chenciner2002,wintner1941analytical}.This is an important reason for us only to consider $d=2$. In fact, A. Chenciner \cite{Chenciner2002} and Zhang-Zhou \cite{zhang2004nonplanar} had proved that the minimizer of Lagrangian action among (anti)symmetric loops for the spatial $N$-body($N\geq4$) problem is a collision-free non-planar solution.\\

\textbf{Notations.} Let
$\mathcal{S}={\{q\in H^1(\mathbb{T}, (\mathbb{R}^2)^N):\int_{\mathbb{T}}{ q(t) dt}=0}\}$.
Let $[x]$ denote the unique integer such that $x-1<[x]\leq x$ for any real $x$. The difference $x-[x]$ is written as $\{x\}$ and satisfies $0\leq\{x\}<1$. \\

First of all, we need a  famous arithmetic fact which belongs to Kronecker:

{\bf Lemma 1. } If 1,$\theta_1$, \ldots, $\theta_n$ are linearly independent over the rational field, then the set \{($\{k\theta_1\}$, \ldots, $\{k\theta_n\}$): $k \in \mathbb{N}\} $ are dense in the $n$-dim unite cube $\{(\varphi_1,\ldots, \varphi_n): 0\leq\varphi_i\leq1, i=1,\ldots, n\}$.\\

The main results in this paper are the following theorems:

{\bf Theorem 1.} Given $\theta_1$, \ldots, $\theta_n$ and any $\epsilon>0$, there are infinitely many integers $k\in \mathbb{N}$ such that $\{k\theta_i\}<\epsilon$ or
$\{k\theta_i\}>1-\epsilon$ for every $i=1,\ldots,n$.\\

{\bf Theorem 2.} If $U(q)\equiv const$, where $q=(q_1,\cdots, q_N)$,
 \begin{equation}
q_i(t)=a_i\cos(\frac{2\pi}{T}t)+b_i\sin(\frac{2\pi}{T}t),~~~~~~\forall t\in\mathbb{T}.
\end{equation}
and $a_i, b_i\in\mathbb{R}^d$ for all $i=1,\ldots, N $. Then $q_i(t)(i=1,\ldots, N) $ is is a rigid motion.

{\bf Corollary 1.}  Saari's Conjecture is true if $i$-th point particle has mode of motion  \begin{equation}
q_i(t)=a_i\cos(\frac{2\pi}{T}t)+b_i\sin(\frac{2\pi}{T}t),~~~~~~\forall t\in\mathbb{T}.
\end{equation}
and $a_i, b_i\in\mathbb{R}^d$, for all $i=1,\ldots, N $.

{\bf Corollary 2.}  Saari's Conjecture is true if in a barycentric reference frame the configurations formed by the bodies remain
the  central configurations all the time.

{\bf Remark.}  If Finiteness of Central Configurations is true \cite{hampton2006finiteness,smale1998mathematical,wintner1941analytical}, the proposition is obvious. But we don't need this hypothesis here.\\

{\bf Theorem 3.} The regular solutions of the N-body problem which minimize the functional ${\mathcal{A}}$ in $\mathcal{S}$ are precisely a relative equilibrium solution whose
configuration minimizes the function $IU^2$ in ${\mathbb{R}^2}$.

{\bf Remark.}  Compared with the result of A.Chenciner \cite{Chenciner2002}: For the planar $N$-body problem, a relative equilibrium solution whose
configuration minimizes $I^\frac{1}{2}U$
is always a minimizer of the action among (anti)symmetric
loops; moreover, all minimizers are of this form provided there exists only a finite number
of similitude classes of $N$-body central configurations. For the second part, he could only prove rigorously for 3-body and 4-body problems, since we
know that the Finiteness of Central Configurations have only been proved for 3-body and 4-body problems until now.
\section{The Proof}
\label{sec:1}
~~~~{\bf Proof of Theorem 1:}

If all of $\theta_1$, \ldots, $\theta_n$ are rational, the proposition is obviously right. Hence, without loss of generality, we will suppose that 1,$\theta_1$, \ldots, $\theta_l$($1\leq l\leq n$) are linearly independent over the rational field and $\theta_{l+1}$, \ldots, $\theta_n$ can be spanned by rational linear combination. That is, we have $\theta_i=x_i^0+\sum_{1\leq j\leq l}x_i^j\theta_j$, where $l<i\leq n$ and $x_i^j$ are rational numbers for $0\leq j\leq l$. Let integer $p$ satisfy that all of $px_i^0$ are integers for $l<i$. It is easy to know that 1,$p\theta_1$, \ldots, $p\theta_l$ are still linearly independent over the rational field. Then for any $\delta>0$, there are infinitely many integers $k\in \mathbb{N}$ such that $\{kp\theta_i\}<\delta$ or
$\{kp\theta_i\}>1-\delta$ for every $i=1,\ldots,l$ by the ${\mathbf{Lemma~ 1}}$ in Section 1, and it is easy to know that $\{kp\theta_i\}<C\delta$ or
$\{kp\theta_i\}>1-C\delta$  for some constant $C$ which only depends on $x_i^j$. So for any $\epsilon>0$, there are infinitely many integers $k\in \mathbb{N}$ such that $\{k\theta_i\}<\epsilon$ or
$\{k\theta_i\}>1-\epsilon$ for every $i=1,\ldots,n$.\\

$~~~~~~~~~~~~~~~~~~~~~~~~~~~~~~~~~~~~~~~~~~~~~~~~~~~~~~~~~~~~~~~~~~~~~~~~~~~~~~~~~~~~~~~~~~~~~~~~~~~~~~~~~~~~~~~~\Box$\\

{\bf Proof of Theorem 2:}

Firstly, we represent $U(q(t))$ as Fourier series.
\begin{eqnarray}
U& = &\sum_{1\leq j<k\leq N} \frac{m_jm_k}{|q_j-q_k|}\nonumber\\
& = &\sum_{1\leq j<k\leq N} \frac{m_jm_k}{[|a_j-a_k|^2\cos^2(\frac{2\pi}{T}t)+|b_j-b_k|^2\sin^2(\frac{2\pi}{T}t)+2(a_j-a_k)\cdot(b_j-b_k)\sin(\frac{2\pi}{T}t)\cos(\frac{2\pi}{T}t)]^\frac{1}{2}}\nonumber\\
& = &\sum_{1\leq j<k\leq N} \frac{m_jm_k}{[\frac{|a_j-a_k|^2+|b_j-b_k|^2}{2}+(\frac{|a_j-a_k|^2-|b_j-b_k|^2}{2}) \cos(\frac{4\pi}{T}t)+(a_j-a_k)\cdot(b_j-b_k)\sin(\frac{4\pi}{T}t)]^\frac{1}{2}}\nonumber\\
& = &\sum_{1\leq j<k\leq N} \frac{m_jm_k}{[\frac{|a_j-a_k|^2+|b_j-b_k|^2}{2}+(\frac{|a_j-a_k|^2-|b_j-b_k|^2}{2}) \cos(\frac{4\pi}{T}t)+(a_j-a_k)\cdot(b_j-b_k)\sin(\frac{4\pi}{T}t)]^\frac{1}{2}}\nonumber\\
& = &\sum_{1\leq j<k\leq N} \frac{m_jm_k}{[A_{jk}+B_{jk}\cos(\frac{4\pi}{T}t+\theta_{jk})]^\frac{1}{2}}\nonumber
\end{eqnarray}
where
 \begin{equation}
A_{jk}=\frac{|a_j-a_k|^2+|b_j-b_k|^2}{2}
\end{equation}
\begin{equation}
B_{jk}=[(\frac{|a_j-a_k|^2-|b_j-b_k|^2}{2})^2 +((a_j-a_k)\cdot(b_j-b_k))^2]^\frac{1}{2}
\end{equation}
and $\theta_{jk}$ can be determined when $B_{jk}>0$.In the following, we will prove $B_{jk}=0$ for any $j, k \in \{{1,\ldots,N}\}$.
 It is easy to know that $A_{jk}\geq B_{jk}$, let $C_{jk}=\frac{B_{jk}}{A_{jk}}$, then we have
\begin{equation}
\begin{aligned}
U & = \sum_{1\leq j<k\leq N} \frac{m_jm_k}{A_{jk}^\frac{1}{2}}[1+(-\frac{1}{2})C_{jk}\cos(\frac{4\pi}{T}t+\theta_{jk})
+\ldots+\\
&\frac{(-\frac{1}{2})(-\frac{1}{2}-1)\ldots(-\frac{1}{2}-n+1)}{n!}(C_{jk})^n\cos^n(\frac{4\pi}{T}t+\theta_{jk})+\ldots]\nonumber\\
& = \sum_{1\leq j<k\leq N} \frac{m_jm_k}{A_{jk}^\frac{1}{2}}\{1+(-\frac{1}{2})C_{jk}\frac{\exp\sqrt{-1}(\frac{4\pi}{T}t+\theta_{jk})+\exp-\sqrt{-1}(\frac{4\pi}{T}t+\theta_{jk})}{2}
 +\ldots+\\
 &\frac{(-\frac{1}{2})(-\frac{1}{2}-1)\ldots(-\frac{1}{2}-n+1)}{n!}(C_{jk})^n[\frac{\exp\sqrt{-1}(\frac{4\pi}{T}t+\theta_{jk})+\exp-\sqrt{-1}(\frac{4\pi}{T}t+\theta_{jk})}{2}]^n+\ldots\}\nonumber\\
 &= \sum_{1\leq j<k\leq N} \frac{m_jm_k}{A_{jk}^\frac{1}{2}}[1+(-\frac{1}{2})C_{jk}\frac{\exp\sqrt{-1}(\frac{4\pi}{T}t+\theta_{jk})+\exp-\sqrt{-1}(\frac{4\pi}{T}t+\theta_{jk})}{2} +\ldots+\\
 &\frac{(-\frac{1}{2})(-\frac{1}{2}-1)\ldots(-\frac{1}{2}-n+1)}{n!}(C_{jk})^n\frac{\sum_{0\leq l\leq n}\left(
                                                                                                       \begin{array}{c}
                                                                                                         n \\
                                                                                                         l \\
                                                                                                       \end{array}
                                                                                                     \right)
\exp\sqrt{-1}((\frac{4\pi}{T}t+\theta_{jk})(2l-n))}{2^n} +\ldots]\nonumber\\
&= \sum_{1\leq j<k\leq N} \frac{m_jm_k}{A_{jk}^\frac{1}{2}}\{1+\sum_{1\leq l}\frac{(-\frac{1}{2})(-\frac{1}{2}-1)\ldots(-\frac{1}{2}-2l+1)}{(2l)!}(C_{jk})^{2l}\frac{\left(
                                                                                                       \begin{array}{c}
                                                                                                         2l \\
                                                                                                         l \\
                                                                                                       \end{array}
                                                                                                     \right)}{2^{2l}}+\\
 &\sum_{1\leq n}\exp\sqrt{-1}(\frac{4n\pi}{T}t)[\frac{(-\frac{1}{2})(-\frac{1}{2}-1)\ldots(-\frac{1}{2}-n+1)}{n!}\frac{(C_{jk})^n\exp\sqrt{-1}(n\theta_{jk})}{2^n}+\\
&\frac{(-\frac{1}{2})(-\frac{1}{2}-1)\ldots(-\frac{1}{2}-n-1)}{(n+2)!}\frac{(C_{jk})^{n+2}\left(
                                                                                                       \begin{array}{c}
                                                                                                         n+2 \\
                                                                                                         n+1 \\
                                                                                                       \end{array}
                                                                                                     \right)\exp\sqrt{-1}(n\theta_{jk})}{2^{n+2}}+\ldots] +\\
 &\sum_{1\leq n}\exp\sqrt{-1}(\frac{-4n\pi}{T}t)[\frac{(-\frac{1}{2})(-\frac{1}{2}-1)\ldots(-\frac{1}{2}-n+1)}{n!}\frac{(C_{jk})^n\exp\sqrt{-1}(-n\theta_{jk})}{2^n}+\\
&\frac{(-\frac{1}{2})(-\frac{1}{2}-1)\ldots(-\frac{1}{2}-n-1)}{(n+2)!}\frac{(C_{jk})^{n+2}\left(
                                                                                                       \begin{array}{c}
                                                                                                         n+2 \\
                                                                                                         n+1 \\
                                                                                                       \end{array}
                                                                                                     \right)\exp\sqrt{-1}(-n\theta_{jk})}{2^{n+2}}+\ldots]\}
\end{aligned}
\end{equation}
Since $U\equiv const$, we have
\begin{equation}
\begin{aligned}
&\sum_{1\leq j<k\leq N} \frac{m_jm_k}{A_{jk}^\frac{1}{2}}[\frac{(-\frac{1}{2})(-\frac{1}{2}-1)\ldots(-\frac{1}{2}-n+1)}{n!}\frac{(C_{jk})^n\exp\sqrt{-1}(n\theta_{jk})}{2^n}+\\
&\frac{(-\frac{1}{2})(-\frac{1}{2}-1)\ldots(-\frac{1}{2}-n-1)}{(n+2)!}\frac{(C_{jk})^{n+2}\left(
                                                                                                       \begin{array}{c}
                                                                                                         n+2 \\
                                                                                                         n+1 \\
                                                                                                       \end{array}
                                                                                                     \right)\exp\sqrt{-1}(n\theta_{jk})}{2^{n+2}}+\ldots]=0
\end{aligned}
\end{equation}
\begin{equation}
\begin{aligned}
&\sum_{1\leq j<k\leq N} \frac{m_jm_k}{A_{jk}^\frac{1}{2}}[\frac{(-\frac{1}{2})(-\frac{1}{2}-1)\ldots(-\frac{1}{2}-n+1)}{n!}\frac{(C_{jk})^n\exp-\sqrt{-1}(n\theta_{jk})}{2^n}+\\
&\frac{(-\frac{1}{2})(-\frac{1}{2}-1)\ldots(-\frac{1}{2}-n-1)}{(n+2)!}\frac{(C_{jk})^{n+2}\left(
                                                                                                       \begin{array}{c}
                                                                                                         n+2 \\
                                                                                                         n+1 \\
                                                                                                       \end{array}
                                                                                                     \right)\exp-\sqrt{-1}(n\theta_{jk})}{2^{n+2}}+\ldots]=0
\end{aligned}
\end{equation}
for any $n\geq1$.
Hence we have
\begin{equation}
\sum_{1\leq j<k\leq N}D^{(n)}_{jk}\exp2\pi\sqrt{-1}(n\frac{\theta_{jk}}{2\pi}) =0
\end{equation}
for any $n\geq1$, where
\begin{equation}
D^{(n)}_{jk}= \frac{m_jm_kC_{jk}^n}{A_{jk}^\frac{1}{2}}[1+\frac{(\frac{1}{2}+n)(\frac{1}{2}+n+1)}{(n+1)(n+2)}\frac{(C_{jk})^2\left(
                                                                                                       \begin{array}{c}
                                                                                                         n+2 \\
                                                                                                         n+1 \\
                                                                                                       \end{array}
                                                                                                     \right)}{2^2}
+\ldots]
\end{equation}
We claim that the right side of (16) is convergent. In fact, let
\begin{equation}
\begin{aligned}
f_{jk}&=1+\frac{(\frac{1}{2}+n)(\frac{1}{2}+n+1)}{(n+1)(n+2)}\frac{(C_{jk})^2\left(
                                                                                                       \begin{array}{c}
                                                                                                         n+2 \\
                                                                                                         n+1 \\
                                                                                                       \end{array}
                                                                                                     \right)}{2^2}
+\ldots\nonumber\\
&=1+c_1(C_{jk})^2+c_2(C_{jk})^4+\ldots+c_l(C_{jk})^{2l}+\ldots
\end{aligned}
\end{equation}
where
\begin{equation}
c_l= \frac{(\frac{1}{2}+n)(\frac{1}{2}+n+1)\ldots(2l-1-\frac{1}{2}+n)(2l-\frac{1}{2}+n)}{(n+1)(n+2)\ldots(n+2l-1)(n+2l)}\frac{\left(
                                                                                                       \begin{array}{c}
                                                                                                         n+2l \\
                                                                                                         n+l \\
                                                                                                       \end{array}
                                                                                                     \right)}{2^{2l}}
\end{equation}
Then we have
\begin{equation}
\frac{c_{l+1}}{c_l}=\frac{(2l+\frac{1}{2}+n)(2l+1+\frac{1}{2}+n)}{4(l+1)(l+1+n)}
\end{equation}
\begin{equation}
\lim_{l\rightarrow\infty}\frac{c_{l+1}}{c_l}=1
\end{equation}
Hence the series (16) is convergent when $(C_{jk})^2<1$. Furthermore, we can prove the convergence of the series (16) by using Gauss' text when $(C_{jk})^2=1$. In fact, we have
\begin{equation}
\frac{c_l}{c_{l+1}}=1+\frac{\frac{n+2}{2}}{l}+\beta_l
\end{equation}
where
\begin{equation}
\beta_l=-\frac{2n^2+2n+\frac{3}{4}+\frac{(n+\frac{1}{2})(n+\frac{3}{2})(n+2)}{2l}}{4l^2+2l(n+2)+(n+\frac{1}{2})(n+\frac{3}{2})}
\end{equation}
Since $\frac{n+2}{2}>1$ and $|\beta_l|\sim\frac{c}{l^2}$, where $c$ is a constant. Then it is easy to know that the series (16) is convergent when $C_{jk}^2=1$.\\
From {Theorem 1}, we know there exists some $n$ such that $n\frac{\theta_{jk}}{2\pi}=k_n+\varphi_{jk}$, where $k_n$ is an integer and $-\frac{1}{4}<\varphi_{jk}<\frac{1}{4}$. Since $D^{(n)}_{jk}\geq0$, there must be $D^{(n)}_{jk}=0$ for any $j,k$ by (13). So we have $C_{jk}=0$, $|q_j-q_k| \equiv \sqrt{A_{jk}}$.
 Hence $q_i(t)(i=1,\ldots, N) $ is a rigid motion.\\
$~~~~~~~~~~~~~~~~~~~~~~~~~~~~~~~~~~~~~~~~~~~~~~~~~~~~~~~~~~~~~~~~~~~~~~~~~~~~~~~~~~~~~~~~~~~~~~~~~~~~~~~~~~~~~~~~\Box$

{\bf Remark.} It is easy to know that the same result is still true when the potential function is defined by
$
U(q) = \sum_{i<j} {\frac{m_i m_j }{|q_i - q_j|^\alpha}}
$ for any $\alpha>0$ and if $U(q(t))$ is a trigonometric polynomial when $i$-th point particle has mode of motion  \begin{equation}
q_i(t)=a_i\cos(\frac{2\pi}{T}t)+b_i\sin(\frac{2\pi}{T}t),~~~~~~\forall t\in\mathbb{T}.
\end{equation}
and $a_i, b_i\in\mathbb{R}^d$, for all $i=1,\ldots, N $.\\

{\bf Proof of Corollary 1:}\\
It is well known that Newtonian particle systems of constant
moment of inertia must satisfy that $U$ is constant.\\
$~~~~~~~~~~~~~~~~~~~~~~~~~~~~~~~~~~~~~~~~~~~~~~~~~~~~~~~~~~~~~~~~~~~~~~~~~~~~~~~~~~~~~~~~~~~~~~~~~~~~~~~~~~~~~~~~\Box$\\

{\bf Proof of Corollary 2:}\\
From the conditions of $ \mathbf{Corollary ~2}$, we have
\begin{equation}
m_i \ddot{q}_i = -\lambda m_iq_i.
\end{equation}
where $\lambda=\frac{U(q)}{I(q)}$ is a constant. Then it is easy to know that
\begin{equation}
q_i(t)=a_i\cos(\sqrt{\lambda}t)+b_i\sin(\sqrt{\lambda}t),~~~~~~\forall t\in\mathbb{T}.
\end{equation}
for some $a_i, b_i\in\mathbb{R}^d$, $i=1,\ldots, N $.\\
$~~~~~~~~~~~~~~~~~~~~~~~~~~~~~~~~~~~~~~~~~~~~~~~~~~~~~~~~~~~~~~~~~~~~~~~~~~~~~~~~~~~~~~~~~~~~~~~~~~~~~~~~~~~~~~~~\Box$\\\\

{\bf Proof of Theorem 3:}\\

We have
\begin{eqnarray}
{\mathcal{A}}(q)& = &\int_{\mathbb{T}}{ [\sum_i \frac{1}{2} m_i |\dot{q_i}|^2  + \sum_{i<j}{\frac{m_i m_j}{|q_i - q_j|}}] dt}\nonumber\\
& \geq &\int_{\mathbb{T}}{[(\frac{2\pi}{T})^2\sum_i \frac{1}{2} m_i |{q_i}|^2  + \sum_{i<j}{\frac{m_i m_j}{|q_i - q_j|}}] dt}\nonumber\\
& = &\int_{\mathbb{T}}{[\frac{1}{2}(\frac{2\pi}{T})^2I(q)  + \frac{1}{2}U(q)+\frac{1}{2}U(q)] dt}\nonumber\\
& \geq &3\int_{\mathbb{T}}{[(\frac{1}{2})^3(\frac{2\pi}{T})^2I(q)U^2(q)]^\frac{1}{3} dt}\nonumber\\
& \geq &3[\frac{(inf _{\mathcal{X}_2\setminus\Delta_2}{IU^2})\pi^2}{2}]^\frac{1}{3}T^\frac{1}{3}\nonumber
\end{eqnarray}
then, ${\mathcal{A}}(q)=3[\frac{(inf_{\mathcal{X}_2\setminus\Delta_2}{IU^2})\pi^2}{2}]^\frac{1}{3}T^\frac{1}{3}$ if and only if:\\
${(\textit{i})}.$ there exist $a_i, b_i\in\mathbb{R}^2$, for all $i=1,\ldots, N $, such that
 \begin{equation}
q_i(t)=a_i\cos(\frac{2\pi}{T}t)+b_i\sin(\frac{2\pi}{T}t),~~~~~~\forall t\in\mathbb{T}.
\end{equation}
${(\textit{ii})}.$
$
(\frac{2\pi}{T})^2I(q)=U(q).
$\\
${(\textit{iii})}.$  $q$ minimizes the function $IU^2$.\\
By ${(\textit{ii})}$ and ${(\textit{iii})}$ we know $I(q)\equiv const, U(q)\equiv const$, and $q(t)$ is always a central configuration.
Then
 $q$ is a relative equilibrium solution whose
configuration minimizes the function $IU^2$ by Theorem 2.\\
$~~~~~~~~~~~~~~~~~~~~~~~~~~~~~~~~~~~~~~~~~~~~~~~~~~~~~~~~~~~~~~~~~~~~~~~~~~~~~~~~~~~~~~~~~~~~~~~~~~~~~~~~~~~~~~~~\Box$

{\bf Remark.}
 If the Finiteness of Central Configurations is true, ${(\textit{ii})}$ and ${(\textit{iii})}$ are sufficient to prove {Theorem 3}. But this problem don't need so strong Conjecture, it just need the weaker assumption:  the minimum points of the function $IU^2$ are finite. However, as far as we know there doesn't exist rigorous proof under the weaker assumption. So we prove that Saari's conjecture in the particular case (9) to get over the obstacle.\\\\

\end{document}